\documentclass[conference]{IEEEtran}

\IEEEoverridecommandlockouts
\usepackage{cite}
\usepackage{amsmath,amssymb,amsfonts}
\def\BibTeX{{\rm B\kern-.05em{\sc i\kern-.025em b}\kern-.08em
    T\kern-.1667em\lower.7ex\hbox{E}\kern-.125emX}}
\usepackage{graphicx}
\usepackage{textcomp}
\usepackage{soul}
\usepackage{color}
\usepackage{xcolor}
\usepackage{xspace}
\usepackage{multirow}
\usepackage{colortbl}
\usepackage{graphicx}
\usepackage{array}
\usepackage{threeparttable}
\usepackage{makecell}
\usepackage{algorithmicx}
\usepackage{algorithm}
\usepackage{algpseudocode}
\usepackage{xcolor}
\usepackage{url}
\usepackage{float}

\usepackage{microtype} 
\usepackage{balance}

\newcolumntype{C}[1]{>{\centering\arraybackslash}m{#1}}
\algnewcommand\algorithmicforeach{\textbf{for each}}
\algdef{S}[FOR]{ForEach}[1]{\algorithmicforeach\ #1\ \algorithmicdo}
\algnewcommand\algorithmicinput{\textbf{Input:}}
\algnewcommand\algorithmicoutput{\textbf{Output:}}
\algnewcommand\algorithmicdeps{\textbf{Dependencies:}}
\algnewcommand\Input{\item[\algorithmicinput]}
\algnewcommand\Output{\item[\algorithmicoutput]}
\algnewcommand\Deps{\item[\algorithmicdeps]}

\algdef{SE}[DOWHILE]{Do}{doWhile}{\algorithmicdo}[1]{\algorithmicwhile\ #1}
\algnewcommand\algorithmicelseif{\textbf{else if}}
\def\BibTeX{{\rm B\kern-.05em{\sc i\kern-.025em b}\kern-.08em
    T\kern-.1667em\lower.7ex\hbox{E}\kern-.125emX}}

\newcommand{\tool}{\textsc{CTG}\xspace}
\newcommand{\botium}{\textsc{Botium}\xspace}
\newcommand{\charm}{\textsc{Charm}\xspace}

\newcommand{\numiterations}{15\xspace}

\AtBeginEnvironment{algorithmic}{\fontsize{6}{5.5}\selectfont}
\algrenewcommand\alglinenumber[1]{\scriptsize#1}

\begin{document}
\title{Test Case Generation for Dialogflow Task-Based Chatbots}

\author{
\IEEEauthorblockN{Rocco Gianni Rapisarda}
\IEEEauthorblockA{\textit{Dep. of Informatics, Systems and Communication} \\
\textit{University of Milano-Bicocca}\\
Milan, Italy \\
r.rapisarda2@campus.unimib.it}
\and 
\IEEEauthorblockN{Davide Ginelli}
\IEEEauthorblockA{\textit{Dep. of Informatics, Systems and Communication} \\
\textit{University of Milano-Bicocca}\\
Milan, Italy \\
davide.ginelli@unimib.it}
\and
\IEEEauthorblockN{Diego Clerissi}
\IEEEauthorblockA{\textit{Dep. of Informatics, Bioeng., Robotics and Systems Engineering} \\
\textit{University of Genova}\\
Genova, Italy \\
diego.clerissi@dibris.unige.it}
\and
\IEEEauthorblockN{\hspace{1.9cm}Leonardo Mariani}
\IEEEauthorblockA{\hspace{1.41cm}\textit{Dep. of Informatics, Systems and Communication} \\
\textit{\hspace{1.4cm}University of Milano-Bicocca}\\
\hspace{1.4cm}Milan, Italy \\
\hspace{1.6cm}leonardo.mariani@unimib.it}
}

\maketitle

\begin{abstract}
Chatbots are software typically embedded in Web and Mobile applications designed to assist the user in a plethora of activities, from chit-chatting to task completion. They enable diverse forms of interactions, like text and voice commands. As any software, even chatbots are susceptible to bugs, and their pervasiveness in our lives, as well as the underlying technological advancements, call for tailored quality assurance techniques. 
However, test case generation techniques for conversational chatbots are still limited. 
In this paper, we present Chatbot Test Generator (\tool), an automated testing technique designed for task-based chatbots. We conducted an experiment comparing \tool with state-of-the-art \botium and \charm tools with seven chatbots, observing that the test cases generated by \tool outperformed the competitors, in terms of robustness and effectiveness. 
\end{abstract}

\begin{IEEEkeywords}
Chatbot Testing, Test Automation, Coverage, Flakiness, Mutation Testing
\end{IEEEkeywords}

%

\section{Introduction}\label{sec:introduction}


\emph{Task-based chatbots} are chatbots designed to deliver functionalities through conversations. Notable examples are chatbots for assisting users, booking services, and completing transactions~\cite{adamopoulou2020overview,adamopoulou2020chatbots}. 
In recent years, task-based chatbots (hereafter simply chatbots) have gained popularity thanks to their integration into Web and Mobile applications, and advances in underlying technologies have made them ubiquitous in a multitude of domains, such as e-commerce, booking, tech support, healthcare, and more~\cite{grudin2019chatbots,folstad2017chatbots,shawar2007chatbots,adamopoulou2020overview,adamopoulou2020chatbots}. Further, chatbots have become significant in the context of business companies, as they allow for a reduction in personnel expenses by providing a potentially active 24/7 service. 

As chatbots continue to permeate our daily lives supported by a large number of design platforms, such as Dialogflow~\cite{dialogflow}, Amazon Lex~\cite{amazon-lex}, IBM Watson Assistant~\cite{watson-assistant}, and Rasa~\cite{rasa}, ensuring their reliability has become a critical concern. Unlike regular test case generation approaches that validate software through interactions taking the form of, for instance, API calls, HTTP requests, or clicks on visual elements, chatbots require the generation of actual conversations, that is, natural language sentences that correspond to actual user requests. Further, responses also take the form of natural language sentences that must be interpreted to determine their correctness. It is thus imperative to design ad-hoc techniques that can generate test cases that thoroughly exercise software whose interface is conversational~\cite{deriu2021survey,li2022review}.

The initial effort in test case generation for chatbots was mainly focused on exercising speech recognition capabilities~\cite{iwama2019automated,asyrofi2020crossasr,schonherr2018adversarial,zhang2019life,qin2019imperceptible}, and other non-functional characteristics~\cite{chatbottest}. The generation of test cases (i.e., conversations) to thoroughly validate the functionalities implemented by the chatbot under test required extensive manual intervention~\cite{alexa-test-framework,bespoken,rasa,vasconcelos2017bottester,ruane2018botest,guichard2019assessing,bozic2019testing,bovzic2022ontology}. 
So far, two approaches have targeted automatic test case generation for chatbots: \botium~\cite{botium}, which is a state-of-the-practice tool originally developed by Botium GmbH (now Cyara~\cite{cyara}), and \charm~\cite{bravo2020testing}, which extends \botium with the capability to generate more diverse test cases.  However, both approaches are \emph{not able to thoroughly cover the  conversational input space} of a chatbot and often generate \emph{incorrect test cases}, due to the challenge of predicting the responses that must be produced by tested chatbots. 

To address the need for an automated testing solution capable of generating robust and effective test cases that can be used, this paper presents Chatbot Test Generator (\tool), an automated testing tool designed for Dialogflow task-based chatbots. \tool exploits the tests generated by \botium as seed tests that are systematically \emph{augmented} to produce test cases that cover \emph{alternative conversational paths}. Moreover, \tool  is the first \emph{dynamic test generation approach} for chatbots, that is, it executes the test cases and records the chatbot responses to obtain test cases that can be faithfully re-executed as regression tests, compared to statically generated tests that may include incorrect oracles. Finally, \tool generates test cases with \emph{setup} and \emph{teardown} operations that prepare and clean up the environment, preventing any accidental interference with the generated test cases.
We conducted an experiment comparing \tool with \botium and \charm, state-of-the-art tools, observing that the test cases generated by \tool outperformed those generated by competing approaches, in terms of both robustness and effectiveness. 
In particular, \tool generated 95\% correct test cases, compared to \botium and \charm that generated 61\% and 82\% correct tests, respectively. Moreover, the tests generated by \tool killed the highest number of mutants in five out of the seven chatbots considered in our mutation testing experiment. This result suggests that the tests obtained with \tool can be used to create better regression test suites than using state-of-the-art approaches.


This paper provides the following contributions: (i) it proposes \tool, the first dynamic test case generation tool for Dialogflow task-based chatbots, (ii) it presents a systematic strategy to generate reliable conversations that cover the possible conversational paths, including the usage of entity values, alternative utterances, and set up and tear down of the testing environment, and, (iii) it presents empirical results that show how \tool outperforms the \botium and \charm state-of-the-art tools when generating tests for seven task-based chatbots.

The paper is organized as follows. Section~\ref{sec:chatbots} introduces the key elements of task-based chatbots. Section~\ref{sec:tool} describes \tool. Section~\ref{sec:evaluation} presents our experiment. Section~\ref{sec:related} discusses related work. Then, Section \ref{sec:conclusion} provides final remarks.


\section{Dialogflow Task-Based Chatbots} \label{sec:chatbots}
  
Chatbots, also known as virtual assistants or conversational agents, are software programs designed to interact with human users through textual, vocal, images, or other communication channels~\cite{adamopoulou2020chatbots}. Chatbots can be categorized based on their \emph{operational focus} (i.e., informative, conversational, and task-based), their \emph{context} (i.e., domain-specific or domain-independent), and the \emph{technology and modules} they are composed of (e.g., natural language processors, speech recognition systems, machine learning models)~\cite{adamopoulou2020overview,adamopoulou2020chatbots}.
Task-based chatbots are chatbots designed to efficiently complete specific tasks through user conversations, prioritizing functionality over casual chat. They aim to fulfill a user's intent, such as booking a service or guiding users through procedures. 

Dialogflow~\cite{dialogflow} is a Natural Language Understanding (NLU) platform, part of the Google Cloud Platform, and one of the most popular platforms for multi-language chatbot design~\cite{abdellatif2021comparison}. 
The platform offers a user-friendly Web interface, advanced machine learning and voice recognition modules for NLU, and native support with Google Cloud services. Further, it integrates with existing services (e.g., Slack, Telegram, Messenger) and custom external services, via Cloud Functions for Firebase. Finally, it exposes APIs that enable connections with existing testing tools, such as Botium~\cite{botium}.
In terms of structure, a Dialogflow chatbot is composed of several components encapsulating the main concepts of a conversation~\cite{canizares2022automating,ferdinando2024mutabot}: 
\begin{itemize}
	\item \emph{Intents}: they represent the possible goals of a user that the chatbot considers as tasks to achieve; users express their intent using utterances (e.g., the intent to \emph{order a pizza} can be expressed by the sentence \emph{I would like to order a pizza});
	\item \emph{Actions}: they are the actions that a chatbot may perform to complete a task; actions can be either plain text responses/requests (e.g., asking the question \emph{What pizza would you like to order?}), graphical objects to interact with, or requests invoking external services; 
	\item \emph{Entities}: the data types that can be used in a conversation, either representing simple literals (e.g., \emph{Pepperoni pizza}), regular expressions, or compositions of other entities;
	\item \emph{Flows}: the possible user-bot interactions composing conversations, which can include context data representing the time-bounded memory of the chatbot. 
\end{itemize}

Testing a task-based chatbot requires generating conversations that thoroughly cover the intents, entities, and flows that are defined in a chatbot. 

\section{Chatbots Test Generator}\label{sec:tool}\begin{figure*}[t!]
    \centering
		\includegraphics[trim=0.0cm 0.0cm 0.0cm 0.0cm, clip=true, width=0.7\linewidth]{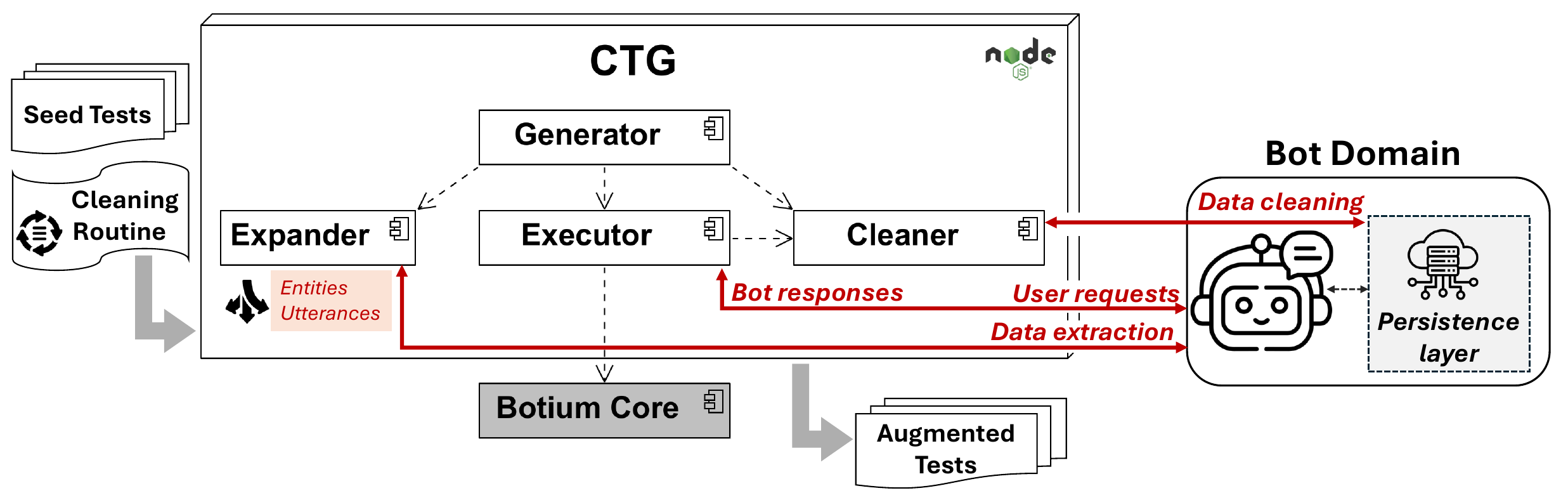}
		\vspace{-1mm}
    \caption{\tool architecture.}
    \label{fig:ctg}
		\vspace{-1mm}
\end{figure*}

Chatbot Test Generator (CTG) is an automated test generator tool for Dialogflow chatbots, developed in Node.js. Its architecture is shown in Figure \ref{fig:ctg}. 

Starting from a set of \botium seed tests that exercise the intents defined in the chatbot under test (e.g., one seed test per intent), \tool generates augmented tests that cover inputs often ignored by test case generation. The seed tests represent the backbone of the conversations explored by \tool. Further, \tool is able to set up and tear down the environment of the chatbot, preventing any side-effect between test executions.

\begin{algorithm}[th!]
\scriptsize
\begin{algorithmic}[1]
\caption{\tool test case generation process.}\label{alg:generator}
\Input{
\\
 $ST$: the seed tests \\
 $CR$: the cleaning routine to set up the connection with the chatbot \\
 $bot$ : the bot under test
}
\Output{
\\
$AT$: the augmented tests
}
\Deps{
\\
$\textbf{function }\textsc{expand}(st, botMsg, bot)$ \Comment{Alg.\ref{alg:expander}}\\
$\textbf{function }\textsc{setUp}(CR)$ \Comment{Alg.\ref{alg:cleaner}}\\
$\textbf{function }\textsc{tearDown}(CR)$ \Comment{Alg.\ref{alg:cleaner}}
}
\\
\Function{GenerateTests}{\textit{$ST, CR, bot$}}
	\State $ AT \gets \emptyset $
	\For{$st$ in $ST$}
		\State $ startMsgs \gets EXP.expand(st, null, bot) $ 
		\For{$i = 0; i < \lvert startMsgs \rvert ; i++$} 
			\State $ at \gets () $
			\State $ at.seed \gets true $
			\State $ AT[st] \gets AT[st] \cup \{at\} $
		\EndFor
		\For{$i = 0; i < \lvert AT[st] \rvert ; i++$} 
			\State $conn \gets CLEANER.setUp(CR)$		
			\If {$AT[st][i].seed$}
				\State $ userMsg \gets startMsgs[i] $ 
				\Do
					\If {$\lvert userMsg \rvert = 1$} 
						\State $ AT[st][i].addUserMsg(userMsg) $
						\State $ botMsg \gets EXEC.sendMsg(userMsg, conn) $
						\State $ AT[st][i].addBotMsg(botMsg) $
						\State $ userMsg \gets EXP.expand(st, botMsg, bot)$ 
					\Else 
						\For{$j = 1; j < \lvert userMsg \rvert ; j++$}
							\State $at^\prime \gets () $
							\State $ at^\prime.addUserMsg(AT[st][i].getUserMsgs()) $ 
							\State $ at^\prime.addUserMsg(userMsg[j]) $ 
							\State $ at^\prime.seed \gets false $ 
							\State $ AT[st] \gets AT[st] \cup \{at^\prime\} $
						\EndFor
						\State $ AT[st][i].addUserMsg(userMsg[0]) $
						\State $ botMsg \gets EXEC.sendMsg(userMsg[0], conn) $
						\State $ AT[st][i].addBotMsg(botMsg) $
						\State $ userMsg \gets EXP.expand(st, botMsg, bot)$					
					\EndIf
				\doWhile{$userMsg \neq null$}
			\Else 
				\State $ userMsgs \gets AT[st][i].getUserMsgs() $
				\For{$j = 0; j < \lvert userMsgs \rvert ; j++$}
					\State $ botMsg \gets EXEC.sendMsg(userMsgs[j], conn) $
					\State $ AT[st][i].addBotMsg(botMsg) $
				\EndFor
			\EndIf
			\State $conn \gets CLEANER.tearDown(CR)$	
		\EndFor
	\EndFor
	\State \Return $ AT $
\EndFunction
\end{algorithmic}
\end{algorithm}

Unlike other approaches that generate tests for chatbots using static analysis (i.e., without executing the chatbot), \tool incrementally generates test cases by sending user messages and recording bot responses at runtime, thus producing test cases that resemble actual conversations, being more reliable for regression testing than statically generated tests.
\tool consists of four key components. The \textsc{Generator} orchestrates the entire test generation process. The \textsc{Expander} retrieves alternative utterances and entity values that can be used as replacements for those present in the seed test cases, then it unrolls the user-bot interactions to explore additional conversational paths based on the extracted alternative values, potentially exercising unexplored intents. The \textsc{Executor} runs the generated tests, using \botium connector module to interact with the chatbot deployed on Dialogflow, and records the conversation (i.e., user requests and bot responses). 
The \textsc{Cleaner} is configured according to a \emph{cleaning routine} defined by the tester to set up and tear down the environment, including the persistence layer of the chatbot under test, thus favoring testing in a neutral environment.  
In the following, we present the main algorithms of \tool in detail.

\subsection{Generator}

Algorithm \ref{alg:generator} shows the \emph{generateTests} function implemented by the \textsc{Generator}. The function takes in input the seed tests $ST$, used as a basis for the generation process, the cleaning routine $CR$, used to configure the \textsc{Cleaner} with the logic to connect to the chatbot and to set up and tear down its back-end, and the reference to the chatbot under test $bot$, used to extract the alternative utterances and entity values from the chatbot, and returns the collection $AT$ of augmented tests for each seed test  (e.g., $AT[st_i][j]$ is the $j^{th}$ augmented test for seed test $i$). 

The seed tests $ST$ must cover the conversational flows that should be used to generate alternative conversations. Although seed tests could be obtained in multiple ways, including manually implementing them, in our assessment we used \botium to generate seed tests, since \botium can automatically generate a set of test cases that exercise the main conversational flow defined in a chatbot. 

The function \emph{generateTests} iterates over each seed test case to generate corresponding alternative test cases (lines 11-51). The \textsc{Generator} starts by using the \textsc{Expander} to retrieve all alternative messages that a user can use to start a conversation according to the set of salutation messages expected by the chatbot (line 12). To perform such a task, the \textsc{Expander} accesses the chatbot structural files, extracting the alternative utterances and entity values. 
For each message that can be used to start a conversation, a new empty augmented test case is created and added to the output set (lines 13-17). 
Each new test case is flagged as ``seed'' (line 15), since the algorithm discriminates tests that cover the seed conversational flow 
from the alternatives that may originate by the expansion of the input data in future interactions (e.g., after the greetings, the conversation may take a new path if the user orders a pizza instead of a hamburger).

For each new test case (lines 18-50), the connection to the chatbot is established through the \textsc{Cleaner} (line 19). 
If the test is flagged as ``seed'' (line 20), a proper salutation message is used to initiate the test from the available ones (line 21). Then, the conversation is iteratively built and executed until no more messages can be sent to the chatbot (lines 22-41). Within each iteration, if the \textsc{Expander} has not identified any alternative message to be sent, the only possible message is sent to the chatbot using the \textsc{Executor}, which exploits Botium APIs to communicate with the bot deployed on Dialogflow. The bot response is collected and encoded as expectation in the test, and the next user message to be sent is retrieved from the alternative messages identified by the \textsc{Expander} (lines 23-27). In case, instead, there are multiple alternatives for the user message to be sent, for each alternative, a new augmented test case is created (lines 29-35): every alternative test inherits the initial part of the conversation already established within the seed test from which the new test is created (line 31) and such test is added to the output set of augmented tests associated with the seed test (line 34). Then, the user-bot interaction proceeds for the seed test (lines 36-39), using the first user message among the alternatives that, by construction, indicates a seed test (i.e., $userMsg[0]$, line 36).

Once an augmented test has reached the end of the execution, the alternative augmented test cases that are pending are retrieved and completed using the same strategy (lines 42-48). 
Eventually, at the end of each test case interaction with the bot, the \textsc{Cleaner} is called to restore the chatbot to its original state (line 49). 
The aforementioned procedure is repeated until no more seed tests are present, and finally the whole set of augmented tests is returned (line 52). 

\subsection{Expander}
Algorithm \ref{alg:expander} shows the \emph{expand} function implemented by the \textsc{Expander}. The function takes in input a test case $t$, a bot message $botMsg$ occurring in the test, the reference to the chatbot $bot$, and returns all the alternative user responses $alt$ that can be sent to the chatbot for the message $botMsg$. 

\begin{figure*}[t!]
    \centering
		\includegraphics[trim=0.0cm 0.0cm 0.0cm 0.0cm, clip=true, width=0.6\linewidth]{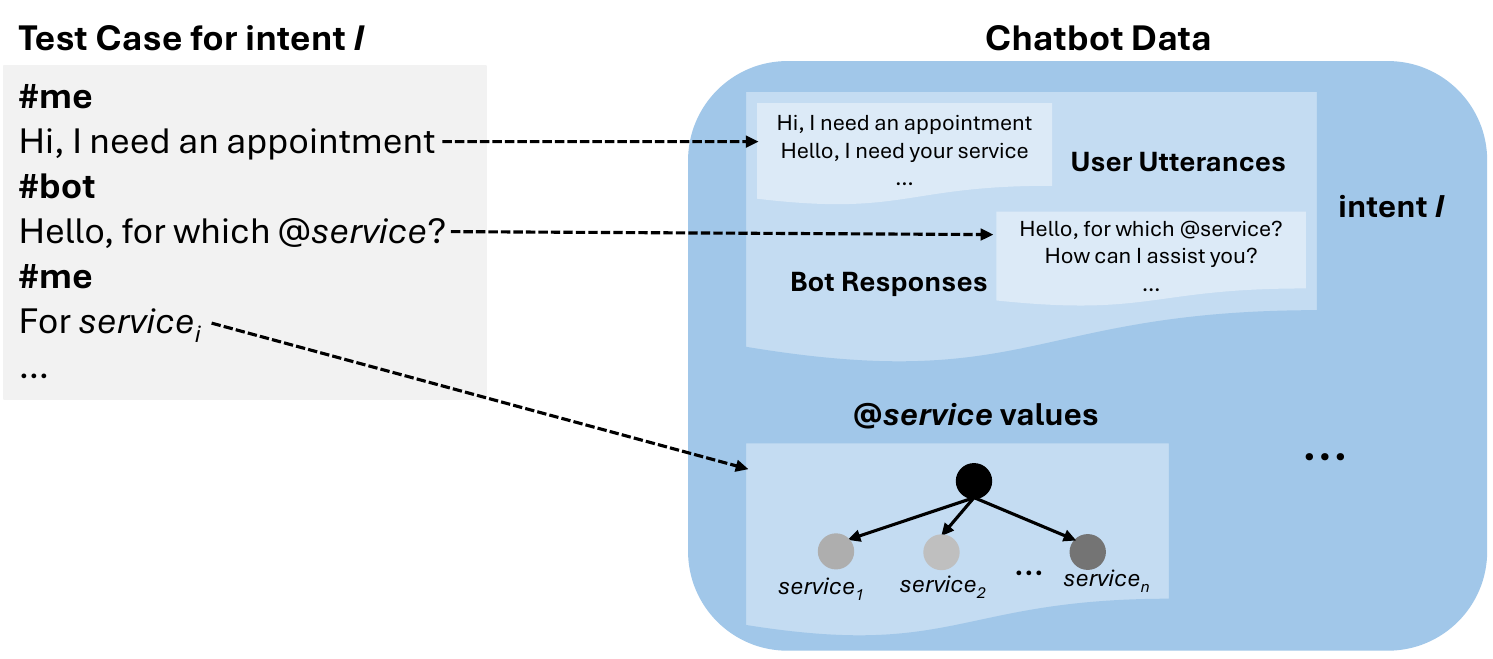}
		\vspace{-1mm}
    \caption{Overview of a \botium test case and Dialogflow chatbot data.}
    \label{fig:test}
    \vspace{-1mm}
\end{figure*}

To perform its tasks, the \textsc{Expander} accesses the Dialogflow data defined for the bot and exploits the \botium test case structure, outlined in the example in Figure~\ref{fig:test}. In \botium, a test case is a conversation that alternates user sentences, occurring after the tag \texttt{\#me}, and bot sentences, occurring after the tag \texttt{\#bot}. For each chatbot intent, defined by user utterances and corresponding chatbot responses (\emph{User Utterances} and \emph{Bot Responses} sections in Figure~\ref{fig:test}), used internally by the chatbot to train a model to recognize sentences flexibly, \botium exploits the dataset and builds the seed test cases as finite combined sequences of user-bot interactions, where the bot messages can be used as oracles to compare with actual responses during test execution.

When a sentence refers to an entity, the bot identifies it with the \texttt{@} symbol (e.g., in the example, the bot is asking for the service the user wants). The chatbot has a number of entity values that can be accepted for each entity. During the conversation, the user must choose one of them to produce a response that is understandable by the bot. The \textsc{Expander} accesses this information to find any alternative utterance and entity value that can be used in the conversation. 

If the bot message to answer is assigned with the value \texttt{null} (lines 7-9), this means that the conversation has just started. The user message for which the alternatives must be found is the one starting the test case (e.g., the first \texttt{\#me} message of Figure~\ref{fig:test}); then, the \textsc{Expander} can extract the alternative sentences by accessing the user utterances file associated with that message (\emph{User Utterances} section in Figure~\ref{fig:test}). If, instead, the bot message is a prompt referring to one or more required entity values (lines 10-11), where an entity is determined by the presence of the ``@'' symbol, 
the alternative messages are obtained by combining the entity values extracted for each entity from the chatbot (\emph{@service values} section in Figure~\ref{fig:test}); the logic of entity value extraction and combination is encapsulated by the method \emph{getEntityValues}, not shown for space reasons. Else, if the bot message is a simple plain text response (lines 12-15), the next user message is retrieved from the test case (line 13), and the alternative user responses are extracted from the utterances file related to the user message (line 14).
Finally, the set of alternatives messages discovered is returned to the \textsc{Generator} (line 16), at its disposal for the test generation process.

\begin{algorithm}[b]
\scriptsize
\begin{algorithmic}[1]
\caption{\tool alternatives expander process.}\label{alg:expander}
\Input{
\\
 $t$: the test case to expand data from \\
 $botMsg$: the bot message used to retrieve the next user message \\
$bot$ : the bot under test
}
\Output{
\\
$alt$: the alternative responses (e.g., entities, utterances)
}
\\
\Function{expand}{\textit{$t, botMsg, bot$}}
\If {$botMsg = null$}
		\State $userMsg \gets getFirstUserMsg(t)$
		\State $alt \gets getUtterances(userMsg, bot) $
	\ElsIf{$botMsg.contains(``@") $} 
		\State $alt \gets getEntityValues(botMsg, bot) $
	\Else 
		\State $userMsg \gets getNextUserMsg(t, botMsg)$
		\State $alt \gets getUtterances(userMsg, bot) $
	\EndIf
\State \Return $alt$
\EndFunction
\end{algorithmic}
\end{algorithm}

%


\subsection{Cleaner}
Algorithm \ref{alg:cleaner} shows the \emph{setUp} and \emph{tearDown} functions of the \textsc{Cleaner} component. 
The \emph{setUp} function is called at the beginning of the generation process (line 19 from Algorithm \ref{alg:generator}) to establish a connection to the chatbot under test and to its persistence layer; it takes in input the cleaning routine $CR$ to inject to the \textsc{Cleaner} the logic to manage the exact cleaning process for the target chatbot (e.g., how to remove events from a Google Calendar for a calendar managing chatbot), and returns a connection $conn$ to communicate with the chatbot. 
More in detail, the service account configuration pertaining the external service that the chatbot connects to, if any, is built (line 7); a service account could refer, for instance, to the Google API managing Google Calendar, thus requiring data such as private keys and project ids to later connect to such service. Next, the connection to the chatbot is opened and returned (lines 8-9). 
Eventually, the connection pointer and the cleaning routine are used in the \emph{tearDown} function, called at the end of each test generation process (line 49 from Algorithm \ref{alg:generator}), to restore the chatbot to its original state, by querying the service for any persistent item to clean (lines 13-15), such as events from a given calendar; finally, the connection is closed (lines 16-17).

\begin{algorithm}[t]
\scriptsize
\begin{algorithmic}[1]
\caption{\tool cleaning process.}\label{alg:cleaner}
\Input{
\\
 $CR$: cleaning routine.
}
\Output{
\\
$conn$: the connection session to the chatbot
}
\Deps{
\\
$\textbf{function }\textsc{clean}(CR)$ \Comment{Extern}\\
$\textbf{function }\textsc{items}()$ \Comment{Extern}\\
}
\Function{SetUp}{\textit{$CR$}}
	\State $serviceAccount \gets buildService(CR)$
	\State $conn \gets connect(serviceAccount, CR)$
	\State \Return $conn$
\EndFunction
\\
\Function{TearDown}{\textit{$CR$}}
	\If {$conn \neq null \land \lvert conn.items() \rvert \geq 1$}
		\State $conn.clean(CR)$
	\EndIf
	\State $conn \gets null$
	\State \Return $null$
\EndFunction
\end{algorithmic}
\end{algorithm}

\section{Empirical Evaluation}\label{sec:evaluation}
  
To study the effectiveness of Chatbot Test Generator (\tool) we considered the following research questions:
\begin{itemize}
	\item \textbf{RQ1-Correctness}: \emph{How often does \tool generate semantically correct tests?} This research question studies the reliability of the generated test cases, considering the rate of wrong tests (i.e., tests that may fail even for passing executions) that are generated.
	\item \textbf{RQ2-Coverage}: \emph{How thoroughly can  \tool exercise conversations?} This research question studies how thoroughly the generated tests exercise the intents and entities implemented in chatbots.
	\item \textbf{RQ3-Mutations}: \emph{What is the defect detection capability of the tests generated by \tool?} This research question studies how well test cases can reveal faults injected in conversations implemented in task-based chatbots.
\end{itemize}

The three research questions are investigated by comparing the test cases generated by \tool with the test cases generated by the baseline approach \botium~\cite{botium} and the state-of-the-art approach \charm~\cite{bravo2020testing}, with seven task-based chatbots selected from open-source third-party repositories. In the rest of this section, we describe the subject chatbots used for the evaluation (Section~\ref{sec:subjects}), the competing techniques (Section~\ref{sec:techniques}), the experimental procedure defined to answer the research questions (Section~\ref{sec:experiments}), the empirical results obtained for the three research questions (Sections~\ref{sec:rq1},~\ref{sec:rq2}, and~\ref{sec:rq3}), and the threats to validity (Section~\ref{sec:threats}).

\subsection{Subject Chatbots} \label{sec:subjects}

\begin{table}[t!]
\caption{Subject Chatbots.}
\vspace{-1mm}
\label{tab:subjects}
\centering
\renewcommand{\arraystretch}{0.9}
\begin{tabular}{|C{2.9cm}|C{1.6cm}|C{1.2cm}|C{1.3cm}|}
    \hline
    \textbf{Name} & \textbf{Domain} & \textbf{\# Intents} & \textbf{\# Entities (\# Values)} \\
    \hline
     E-Commerce\tnote & Shopping & 11 & 0 (0) \\ \hline
     Room Reservation & Hotels & 6 & 1 (3) \\ \hline
     Weather Forecast & Weather  & 4 & 0 (0)\\ \hline
		 Currency Converter & Currencies  & 3 & 0 (0)\\ \hline
     Temperature Converter & Temperatures  & 5 & 2 (12)\\ \hline
     Appointments Scheduler & Cars  & 3 & 1 (10)\\ \hline
     News & Info & 3 & 1 (9)\\ 
    \hline
\end{tabular}
\vspace{-3mm}
\end{table}

To investigate the research questions, we selected seven Dialogflow chatbots from the ASYM0B repository of open-source third-party chatbots~\cite{original-chatbots-repo}, which has already been used in previous studies~\cite{ferdinando2024mutabot,canizares2022automating,bravo2020testing}. The selected chatbots are representative of different domains (e.g., hotels booking, weather forecasting), and designed to cover non-trivial conversational aspects (e.g., nested intents, input contexts, and back-end components to interact with external services).
Table~\ref{tab:subjects} reports the name, the domain, and the conversational size of each selected chatbot (i.e., the number of intents, custom entities, and entity values). 
 
 Since the \emph{Weather Forecast} and \emph{News} chatbots did not include a configuration file for the back-end component, we had to define one ourselves to integrate with free API services alternative to the original, unavailable, ones. We specifically used the OpenWeather and News APIs. 
 We carefully inspected and tested the modified chatbots to ensure they preserved their integrity, conversational structure, and coherence. 

\subsection{Competing Techniques} \label{sec:techniques}
We selected the \botium\cite{botium} baseline and the \charm\cite{bravo2020testing} state-of-the-art approaches to compare with \tool.
\botium~\cite{botium} is an automated quality assurance framework for chatbot testing designed by Botium GmbH (now Cyara~\cite{cyara}), widely used in both industrial and academic contexts~\cite{bravo2020testing,li2022review}. It offers support for generating and executing test cases on various chatbot design platforms, including Dialogflow~\cite{dialogflow}, Amazon Lex~\cite{amazon-lex} and Rasa\cite{rasa}. It represents the state-of-the-practice approach for test case generation for chatbots. 
\charm~\cite{bravo2020testing} is an open-source testing tool for Dialogflow chatbots implemented in Python. Similarly to \tool, it uses \botium as a back-end for test generation and execution. \charm extends the test case generation capabilities of \botium with mutations over utterances of seed test cases, such as language back-translations and synonym substitutions. 

\subsection{Experimental Procedure} \label{sec:experiments}


To address \textbf{RQ1-Correctness}, we studied the rate of errors present in the tests generated by \tool, \botium and \charm. We separately deployed each subject chatbot on Dialogflow. Then, we used the testing techniques to automatically generate test cases for each chatbot, configuring each tool to use the \emph{multistepconvos} \botium feature, which causes the generation of test cases that tentatively cover full user-bot conversations, according to the conversations defined in the tested chatbots. 
As seed tests required by the \tool generation phase (Algorithm \ref{alg:generator}), we set one \botium seed test per intent.
To assess the correctness of the tests, we executed each test \numiterations times. We then manually inspected the failures to determine the presence of \emph{semantically wrong} test cases, due to malformed utterances, wrong oracles, incoherent interactions, or \emph{flaky interactions\footnote{A flaky test is a test that passes and fails periodically without any code
changes~\cite{zheng2021research}.}}. The remaining tests were deemed as \emph{correct}.

To address \textbf{RQ2-Coverage}, we executed the correct test cases resulting from RQ1 and collected the set of intents and entities exercised by each test. To obtain this information, we run the tests in verbose mode and implemented a script to parse the collected logs, extracting the information about the executed elements of the conversation. We derive two main indicators of the capability of the tests to exercise conversations: \textit{intent coverage} and \textit{entity coverage}~\cite{canizares2024coverage}. Intent coverage is the percentage of intents in the chatbots that are exercised by the test cases, and measures the capability of the tests to cover at least once each user's intent implemented in the chatbot. Entity coverage measures the percentage of actual entity values that are exercised in the tests in comparison to the set of entity values that the chatbot is designed to recognize. For instance, the \emph{Appointments Scheduler} chatbot could be used to schedule appointments about \emph{driver license} or \emph{vehicle registration} (these are possible values for the entity type \emph{AppointmentType}); thus, it indicates how extensive the conversation has been exercised in terms of the kind of values used.

To address \textbf{RQ3-Mutations}, we injected mutants into subject chatbots using Mutabot~\cite{ferdinando2024mutabot}, a mutation testing tool for Dialogflow conversational chatbots. Mutabot generates mutations at multiple levels, including structural changing affecting a conversation (e.g., removing an intent from a chatbot), 
 intents (e.g., flagging an intent as fallback\footnote{A fallback intent is designed to respond to the user requests that the chatbot could not understand.} or changing its priority\footnote{The intent priority determines the order of activation when multiple intents compete for a user request.}), and entities (e.g., removing or changing entity definitions). 
In our evaluation, we applied the following operators affecting the main conversational characteristics of the selected chatbots: 1) intent removal, 2) entity removal, 3) intent parameter removal, 4) intent priority change, 5) intent flagging as fallback, 6) entity renaming, and 7) entity value change.
After generating the mutants, we manually inspected the resulting chatbots to detect and discard the \emph{equivalent}\footnote{An equivalent mutant is a mutant that cannot be killed by any test case~\cite{jia2010analysis}.} mutants, which we could determine by comparing the original and the mutated code elements, and detecting those that had no impact on the behavior of the chatbot (e.g., changing the priority of an intent when there are no other intents to compete with). We independently deployed the mutated chatbots on Dialogflow, and tested them with the test suites produced for RQ1, measuring the rate of mutants killed by each technique. 
To avoid any side effect and guarantee replicability and fairness in the results, we reset the state of the chatbots after each run. 

The resources needed to replicate our experiments, including tools, chatbots, and generated tests, are publicly available at \url{https://gitlab.com/ctg-experiment1/ctg-experiment}.


\subsection{Results for RQ1-Correctness} \label{sec:rq1}

\begin{table}[ht!]
\caption{Generated test cases for each chatbot and technique.}
\vspace{-1mm}
\centering
\setlength{\tabcolsep}{8pt}
\renewcommand{\arraystretch}{0.9}
\footnotesize
\begin{tabular}{|l|c|c|c|}
\hline
\multicolumn{1}{|c|}{\textbf{Chatbot}} & \multicolumn{1}{c|}{\textbf{\botium}} & \multicolumn{1}{c|}{\textbf{\charm}} & \multicolumn{1}{c|}{\textbf{\tool}} \\ \hline
E-Commerce                & 196 & 20 & 196 \\ \hline
Room Reservation          & 233 & 21 & 233 \\ \hline
Weather Forecast         & 11  & 11 & 16  \\ \hline
Currency Converter       & 39  & 17 & 39  \\ \hline
Temperature Converter    & 15  & 15  & 19  \\ \hline
Appointments Scheduler & 21  & 12 & 24  \\ \hline
News                     & 10  & 10 & 19  \\ \hline
\rowcolor{gray!30}
\textbf{Total} & \multicolumn{1}{c|}{\textbf{525}} & \multicolumn{1}{c|}{\textbf{106}} & \textbf{546} \\ 
\hline
\end{tabular}
\label{tab:generated_tests}
\vspace{-4mm}
\end{table}

Table~\ref{tab:generated_tests} lists the number of test cases generated for each chatbot and technique. Overall, \charm generated a smaller set of test cases (106 test cases) compared to \botium and \tool, since it often failed at generating tests for intents whose execution is constrained, such as nested intents that can be exercised only after other intents have been exercised, or intents requiring some input contexts defined by former interactions. The consequence is that \charm misses to exercise several alternative conversations, in particular in the \emph{E-Commerce} and \emph{Room Reservation} chatbots. 

On the other hand, the performance of \botium and \tool is comparable (525 test cases generated by \botium and 546 test cases generated by \tool), since both techniques are able to discover alternative execution paths. However, the \textsc{Expander} component of \tool can also retrieve alternative entity values (see lines 10-11 of Algorithm~\ref{alg:expander}) when the available user utterances are incomplete, and the chatbot must interact with the user to fill the slots with missing entities (e.g., the case of a user requesting for an appointment without specifying the kind of service requested forces that chatbot to produce additional interactions to ask what service is needed to the user). If all the cases are covered by the utterances available in the training data, \tool does not activate the expansion mechanism and the conversations tentatively generated by \tool and \botium are the same. In fact, \tool and \botium, even addressing the same conversational flow,  generate different tests since they rely on different information: \tool generates tests dynamically, based on actual bot responses, while \botium generates tests statically, guessing responses from source files.

\begin{table}[t!]
\centering
\caption{Results of RQ1: Semantically Wrong (SW) and Correct (C) test cases for each chatbot and technique (\%).}
\vspace{-1mm}
\setlength{\tabcolsep}{4.5pt}
\renewcommand{\arraystretch}{0.9} 
\footnotesize
\begin{tabular}{|l|cc|cc|cc|}
\hline
\multicolumn{1}{|c|}{\multirow{2}{*}{\textbf{Chatbot}}} &
  \multicolumn{2}{c|}{\textit{\botium}} &
  \multicolumn{2}{c|}{\textit{\charm}} &
  \multicolumn{2}{c|}{\textit{\tool}} \\ 
	\cline{2-7} 
\multicolumn{1}{|c|}{} &
  \multicolumn{1}{c|}{\textbf{SW}} &
  \textbf{C} &
  \multicolumn{1}{c|}{\textbf{SW}} &
  \textbf{C} &
  \multicolumn{1}{c|}{\textbf{SW}} &
  \textbf{C} \\ 
	\hline
E-Commerce &
  \multicolumn{1}{c|}{0\%} &
  100\% &
  \multicolumn{1}{c|}{0\%} &
  100\% &
  \multicolumn{1}{c|}{0\%} &
  100\% \\ 
	\hline
Room Reservation &
  \multicolumn{1}{c|}{80\%} &
  20\% &
  \multicolumn{1}{c|}{24\%} &
  76\% &
  \multicolumn{1}{c|}{4\%} &
  96\% \\ 
	\hline
Weather Forecast &
  \multicolumn{1}{c|}{36\%} &
  64\% &
  \multicolumn{1}{c|}{18\%} &
  82\% &
  \multicolumn{1}{c|}{40\%} &
  60\% \\ 
	\hline
Currency Converter &
  \multicolumn{1}{c|}{18\%} &
  82\% &
  \multicolumn{1}{c|}{24\%} &
  76\% &
  \multicolumn{1}{c|}{0\%} &
  100\% \\ 
	\hline
Temperature Converter &
  \multicolumn{1}{c|}{27\%} &
  73\% &
  \multicolumn{1}{c|}{53\%} &
  47\% &
  \multicolumn{1}{c|}{0\%} &
  100\% \\ 
	\hline
Appointments Scheduler &
  \multicolumn{1}{c|}{24\%} &
  76\% &
  \multicolumn{1}{c|}{0\%} &
  100\% &
  \multicolumn{1}{c|}{43\%} &
  57\% \\ 
	\hline
News &
  \multicolumn{1}{c|}{0\%} &
  100\% &
  \multicolumn{1}{c|}{0\%} &
  100\% &
  \multicolumn{1}{c|}{0\%} &
  100\% \\ 
	\hline
	\rowcolor{gray!30}
\textbf{Total} &
  \multicolumn{1}{c|}{\textbf{39\%}} &
  \textbf{61\%} &
  \multicolumn{1}{c|}{\textbf{18\%}} &
  \textbf{82\%} &
  \multicolumn{1}{c|}{\textbf{5\%}} &
  \textbf{95\%} \\ \hline
\end{tabular}
\label{tab:rq1}
\vspace{-4mm}
\end{table}

Although the number of test cases generated by \botium and \tool is comparable, the test cases generated by \tool were more robust than those generated by the other techniques. Notice that the test generation timings were comparable and required few seconds across the various tools. Nevertheless, since \tool dynamically generates tests by interacting with the chatbot to record actual responses, the generation time depends on external services that could potentially introduce delays. However, no significant impact was observed for the subject chatbots.

Table~\ref{tab:rq1} reports the results of RQ1. The column \emph{SW} indicates the percentage of semantically wrong test cases that fail when executed, while \emph{C} indicates the percentage of correct test cases that can be reliably executed.
All techniques generated no semantically wrong test cases for \textit{E-Commerce} and \textit{News} chatbots, where dialogues could be reliably extracted from the implementation. Generating reliable test cases was more challenging for the other chatbots. 
\tool generated the highest number of correct test cases for five out of seven chatbots, while \charm outperformed the other techniques in the other two chatbots.
Overall, \botium generated 61\% correct test cases, reporting the lowest performance compared to the other techniques. \charm achieved 82\% correct test cases on average. \tool outperformed the competing techniques, achieving 95\% correct test cases on average.
Note that \charm, in addition to generating a slightly smaller percentage of correct test cases than \tool, generates fewer tests in total.


The reasons for semantically wrong test cases that fail even in case of correct responses are various. We identified four main cases, which we discuss below.

\emph{(1) Oracle Weaknesses:} Chatbots are often designed to provide multiple, semantically equivalent, responses to the same user requests. This can be the source of flakiness, since the expected response specified in a generated test, although correct, can differ with respect to the actual response provided by a chatbot when the test is executed, resulting in spurious failure of an assert statement. This limitation is common to all the compared techniques, since \botium serves as the backbone for test execution, where the oracle is defined by default through text matching between the chatbot expected and actual responses. 
Figure~\ref{fig:flaky_test} shows an example of a flaky test case generated by all the compared techniques for the \emph{Room Reservation} chatbot. The test case simply submits a salutation message and encodes an expectation about the response of the chatbot (column \emph{Expected}). However, the \emph{Room Reservation} chatbot can respond in multiple ways. For example, the actual response could start with \texttt{Good day} rather than \texttt{Hi}, causing a test failure. Implementing a flexible oracle strategy is an open challenge for all compared techniques.

\begin{figure}[t]
\centering
\small
\resizebox{\linewidth}{!}{
\begin{tabular}{lllll}
\hline
\multicolumn{1}{c}{\textbf{Test Case}} &
 &
\begin{tabular}[c]{@{}l@{}}
\textbf{\#me}\\ hello\\ 
\textbf{\#bot}\\ Hi! I'm your room booking bot.\\  I can help you to find a perfect room for your \\ meeting and manage your reservations.
\end{tabular} &
 &
\begin{tabular}[c]{@{}l@{}}
\textbf{\#me}\\ hello\\ 
\textbf{\#bot}\\ \emph{Good day!} I'm your room booking bot.\\ I can help you to find a perfect room for\\ your meeting and manage your reservations.
\end{tabular} \\ 
\hline
\rowcolor{gray!30}
\textbf{} &
 &
\multicolumn{1}{c}{\textbf{Expected}} &
 &
\multicolumn{1}{c}{\textbf{Actual}} \\ 
\hline
\end{tabular}
}
\vspace{-3mm}
\caption{A flaky test case generated by all the techniques.}
\vspace{-6mm}
\label{fig:flaky_test}
\end{figure}

\emph{(2) Dirty Environment:} Flakiness may also occur when the same scenario affecting the chatbot state is exercised multiple times without proper cleanup of the state. While \tool uses the \textsc{Cleaner} component to properly set up and tear down the environment, the test cases generated by both \botium and \charm may interfere with each other. 

\emph{(3) Dynamic Responses:} Both \botium and \charm generate test cases statically by combining the user utterances and the bot responses as they are taken from the chatbot design files, thus they could miss any dynamic response that may depend on the actions executed by the chatbot. 
On the other hand, \tool can collect dynamic responses according to the status of the conversation, as shown in Figure~\ref{fig:dynamic_response} for the \emph{Appointments Scheduler} chatbot, where the final chatbot responses depend on the actual finalization of the user request (e.g., if a slot is available, the appointment will be confirmed).


\begin{figure}[t]
\centering
\resizebox{\linewidth}{!}{
\begin{tabular}{lllll}
\hline
\multicolumn{1}{c}{\textbf{Test Cases}} &
 &
\begin{tabular}[c]{@{}l@{}}
\textbf{\#me}\\ I would like to set an appointment for 3pm \\ on Tuesday\\ 
\textbf{\#bot}\\ What services are you looking to get? \\ DMV offers Driver license and \\ vehicle registration services.\\
\textbf{\#me}\\ Driver License\\
\textbf{\#bot}\\ Let me see if we can fit you in on \\ 2024-05-07 at 15:00! Yes It is fine!\\ 
\end{tabular} &
 &
\begin{tabular}[c]{@{}l@{}}
\textbf{\#me}\\ I would like to set an appointment for 3pm \\ on Tuesday\\ 
\textbf{\#bot}\\ What services are you looking to get? \\ DMV offers Driver license and \\ vehicle registration services.\\
\textbf{\#me}\\ Driver License\\
\textbf{\#bot}\\ I'm sorry, there are no slots available for \\ 2024-05-07 at 15:00. \\
\end{tabular} \\ 
\hline
\rowcolor{gray!30}
\textbf{} &
 &
\multicolumn{1}{c}{\textbf{Scenario \#1}} &
 &
\multicolumn{1}{c}{\textbf{Scenario \#2}} \\ 
\hline
\end{tabular}
}
\vspace{-3mm}
\caption{Test cases generated by \tool capturing dynamic responses.}
\label{fig:dynamic_response}
\vspace{-3mm}
\end{figure}



\emph{(4) Complex Conversational Flow:} The generation strategies implemented by \botium and \charm occasionally neglect the preliminary steps of conversations or skip some necessary interactions. Differently, \tool can address nested intents and interactions, by iteratively building the conversations as long as user requests and bot responses can be exchanged (see lines 22-41 in Algorithm~\ref{alg:generator}). Incorrect user-bot interactions sequences can lead the chatbots to fail understanding the requests, resulting in semantically wrong test cases. 
For example, Figure~\ref{fig:tests_comparison} shows two test cases generated for the \emph{Currency Converter} chatbot. The test case generated by \botium (left side), due to its limited ability to explore nested intents, starts the conversation with a sentence, \texttt{now into Euros}, that is not meaningful alone. In fact, the bot responds with an \texttt{Invalid currency conversion parameters}, failing the test that expects a meaningful response.  In contrast, the test case generated by \tool (right side) implements the entire conversation, including the initial request for a given amount to be converted into a new currency. The test can be repeatedly executed without problems and can be reliably used for regression testing.

\smallskip

To summarize, while oracle weaknesses affect all compared techniques, the other limitations do not affect \tool. 


\begin{figure}[t]
\centering
\small
\resizebox{\linewidth}{!}{
\begin{tabular}{lllll}
\hline
\multicolumn{1}{c}{\textbf{Test Case}} &
 &
\begin{tabular}[c]{@{}l@{}}
\textbf{-} \\ -\\ 
\textbf{-} \\-\\ 
\textbf{\#me}\\now into Euros\\ 
\textbf{\#bot}\\ Invalid currency conversion parameters
\end{tabular} &
 &
\begin{tabular}[c]{@{}l@{}}
\textbf{\#me} \\ Convert 30 Dollars\\ 
\textbf{\#bot} \\What is the currency-to?\\ 
\textbf{\#me}\\now into Euros\\ 
\textbf{\#bot}\\ At the moment 30USD are 27.642EUR       
\end{tabular} \\ 
\hline
\rowcolor{gray!30}
\textbf{} &
 &
\multicolumn{1}{c}{\textbf{\botium}} &
 &
\multicolumn{1}{c}{\textbf{\tool}} \\ 
\hline
\end{tabular}
}
\vspace{-3mm}
\caption{A same test case comparison generated by \botium and \tool.}
\vspace{-4mm}
\label{fig:tests_comparison}
\end{figure}

\subsection{Results for RQ2-Coverage} \label{sec:rq2}

Table \ref{tab:rq2} reports intent and entity coverage achieved for each chatbot and testing technique.

\begin{table}[t!]
\caption{Results of RQ2: Intent and Entity coverage.}
\vspace{-2mm}
\centering
\setlength{\tabcolsep}{1.0pt}
\renewcommand{\arraystretch}{0.8}
\begin{tabular}{|l|cc|cc|cc|}
\hline
\multicolumn{1}{|c|}{\multirow{2}{*}{\textbf{Chatbot}}} &
\multicolumn{2}{c|}{\textit{\botium}} &
\multicolumn{2}{c|}{\textit{\charm}} &
\multicolumn{2}{c|}{\textit{\tool}} \\ \cline{2-7} 
\multicolumn{1}{|c|}{} &
\multicolumn{1}{c|}{\textbf{Intents}} &  \textbf{Entities} &
\multicolumn{1}{c|}{\textbf{Intents}} &  \textbf{Entities} &
\multicolumn{1}{c|}{\textbf{Intents}} &  \textbf{Entities} \\ 
\hline
E-Commerce 
& \multicolumn{1}{c|}{91\%} & -
& \multicolumn{1}{c|}{36\%} & - 
& \multicolumn{1}{c|}{91\%} & - \\ 
\hline
Room Reservation 
& \multicolumn{1}{c|}{83\%} & 100\% 
& \multicolumn{1}{c|}{50\%} & 0\%
& \multicolumn{1}{c|}{83\%} & 100\% \\ 
\hline
Weather Forecast 
& \multicolumn{1}{c|}{50\%} & - 
& \multicolumn{1}{c|}{50\%} & - 
& \multicolumn{1}{c|}{50\%} & - \\ 
\hline
Currency Converter
& \multicolumn{1}{c|}{66\%} & - 
& \multicolumn{1}{c|}{66\%} & - 
& \multicolumn{1}{c|}{66\%} & - \\ 
\hline
Temperature Converter 
& \multicolumn{1}{c|}{60\%} & 33\%
& \multicolumn{1}{c|}{60\%} & 50\%
& \multicolumn{1}{c|}{60\%} & 50\% \\ 
\hline
Appointments Scheduler 
& \multicolumn{1}{c|}{66\%} & 10\%
& \multicolumn{1}{c|}{66\%} & 20\%
& \multicolumn{1}{c|}{66\%} & 20\%\\ 
\hline
News 
& \multicolumn{1}{c|}{66\%} & 11\%
& \multicolumn{1}{c|}{66\%} & 11\% 
& \multicolumn{1}{c|}{66\%} & 56\% \\ 
\hline
\rowcolor{gray!30}
\textbf{Total}  
& \multicolumn{1}{c|}{\textbf{74\%}} & \textbf{26\%} 
& \multicolumn{1}{c|}{\textbf{51\%}} & \textbf{26\%}
& \multicolumn{1}{c|}{\textbf{74\%}} &\textbf{47\%} \\ 
\hline
\end{tabular}
\label{tab:rq2}
\vspace{-4mm}
\end{table}

Concerning intent coverage, all the techniques performed the same with five chatbots: \emph{Currency Converter} (66\%), \emph{Appointments Scheduler} (66\%), \emph{News} (66\%), \emph{Temperature Converter} (60\%), and \emph{Weather Forecast} (50\%). 
Additionally, \botium and \tool both covered all intents but one in \emph{Room Reservation} (83\%) and in \emph{E-Commerce} (91\%), while \charm performed significantly worse in \emph{Room Reservation} (50\%) and \emph{E-Commerce} (36\%), due to the lower number of generated test cases (see again Table~\ref{tab:generated_tests}). 
Overall, \tool and \botium scored the same (74\% of intents covered on average), followed by \charm with 51\%.

Concerning entity coverage, \botium and \tool cover all entities in every chatbot for at least one value. Instead, \charm does not cover the entity in \emph{Room Reservation}, as it failed in generating tests about intents that use such entity. On the other hand, \botium misses more entity values than \charm and \tool in \emph{Temperature Converter} (33\% vs 50\%) and in \emph{Appointments Scheduler} (10\% vs 20\%), whereas it scores the same as \charm in \emph{News} (11\%), making them to tie overall (26\% both). \tool covers 47\% of entity values, scoring the same or better than the other techniques in all cases. 

The main reasons of uncovered intents and entities lie underneath the core functionality of test generation, that is shared among all the techniques. Since the testing tools generate test cases from the training data that reflect the positive usages of the chatbots, they often fail at exercising the fallback intents that are designed to manage the negative scenarios which may originate from realistic but wrong interactions (e.g., ask to rephrase/complete a message received with a badly formatted data)~\cite{ferdinando2024mutabot}. 
Further, some intents may require specific pre-requisite to activate them, such as the contexts data that must be provided in input from past interactions. Among the testing tools, \charm is the one that suffered the most, as observed especially for the \emph{E-Commerce} and the \emph{Room Reservation} chatbots, affecting both test generation and coverage.

Nevertheless, the number of test cases does not necessarily reflect the capability of a technique to cover intents and entities.
For example, we observed cases where radically different number of test cases resulted in the same intent and entity coverage, such as in \emph{Currency Converter} chatbot, where all techniques scored 66\% of intents covered, although \botium and \tool generated 39 test cases while \charm generated 17 only. 
This suggests that the generated tests are sometimes equivalent variations of other existing tests~\cite{canizares2024coverage}, for instance, having a user utterance replaced with another one producing the same effect (e.g., \emph{Convert 10 Pounds to Dollars} versus \emph{How much is 10 Pounds in Dollars?}). 


In a nutshell, all the techniques performed better in covering intents rather than entities, with \tool providing the best combined coverage. Indeed, the achieved intent and entity coverage can still be significantly improved.

\subsection{Results for RQ3-Mutants} \label{sec:rq3}

To assess the fault revealing capability of the compared approaches, we measured the mutation score. To compute this score, we manually inspected the generated mutants and discarded the equivalent ones: we discarded six equivalents from \emph{Room Reservation}, one from \emph{Weather Forecast}, and one from \emph{Currency Converter}. Table~\ref{tab:rq3} reports the mutation score obtained by each technique. 
In five chatbots out of seven (i.e., \emph{Room Reservation}, \emph{Temperature Converter}, \emph{Appointments Scheduler}, \emph{Currency Converter} and \emph{News}) \tool performs better or the same than the other techniques.
\charm performs the same or better than the other approaches in three cases, and \botium performs the same or better than other approaches in one case. Overall, \tool outperformed the competing approaches, killing 57\% of the mutants, with \botium and \charm killing 47\% and 45\%, respectively.



\begin{table}[t!]
\caption{Results of RQ3: Mutants Killed/Total.}
\vspace{-2mm}
\centering
\setlength{\tabcolsep}{5.5pt} 
\renewcommand{\arraystretch}{0.8}
\footnotesize
\begin{tabular}{|l|c|c|c|}
\hline
\multicolumn{1}{|c|}{\multirow{2}{*}{\textbf{Chatbot}}} & \multicolumn{3}{c|}{\textbf{\# Killed / \# Generated (\% Killed)}} \\
\cline{2-4}
& \botium & \charm & \tool \\
\hline
{E-Commerce} & 10/22 (46\%) & \textbf{11/22 (50\%)} & 10/22 (46\%) \\ 
{Room Reservation} & 9/16 (56\%) & 7/16 (44\%) & \textbf{11/16 (69\%)} \\
{Weather Forecast} & 5/13 (38\%) & \textbf{7/13 (54\%)} & 6/13 (46\%) \\
{Currency Converter} & \textbf{3/6 (50\%)} & \textbf{3/6 (50\%)} & \textbf{3/6 (50\%)} \\
{Temperature Converter} & 6/16 (38\%) & 6/16 (38\%) & \textbf{8/16 (50\%)}  \\
{Appointments Scheduler} & 5/12 (42\%) & 4/12 (33\%)  & \textbf{7/12 (58\%)} \\
{News} & 8/12 (67\%) & 6/12 (50\%) & \textbf{10/12 (83\%)}  \\
\hline
\rowcolor{gray!30}
\textbf{Total} & 46/97 (47\%) & 44/97 (45\%) & \textbf{55/97 (57\%)}  \\
\hline
\end{tabular}
\label{tab:rq3}
\vspace{-4mm}
\end{table}



In general, mutants that affected intents and entities (e.g., remove an entity, or remove a required parameter referring to an entity from a user utterance) were easier to kill by \tool, because of its capability of generating test cases that can record the actual chatbot responses and use them as oracles, rather than building static conversations unable to track such details, as well as being able to exercise entities more systematically. For instance, \emph{News} chatbot presents a conversational scenario in which the user has to specify the \emph{topic} of a news; if such parameter is missing, once a test exercises such scenario, the mutant will likely activate the fallback intent as being unable to understand the request prompted by the test, thus getting detected.
On the other hand, some mutations, as reported in previous studies~\cite{ferdinando2024mutabot}, are complex to detect by any technique, such as those affecting negative scenarios (e.g., turning an ordinary intent into a fallback intent), as they are generally neglected by the test generation process. This result confirms that more research is needed in test case generation for task-based chatbots.



\subsection{Threats to Validity} \label{sec:threats}
An internal threat to the validity of this study depends on the manual intervention to deploy the mutated chatbots on Dialogflow, to detect the semantically wrong test cases, and the inspection of the mutants (e.g., to detect equivalent mutants). We mitigated this issue by having two authors carefully inspect the produced artifacts and discuss the result of the inspection until they reached agreement. 
An external threat to validity concerns the generalizability of the results. Although our findings are not final, they provide an interesting picture of the state of the art in testing task-based chatbots. The assessment considers chatbots from different domains, using third-party software also involved in other experiments~\cite{ferdinando2024mutabot,canizares2022automating,bravo2020testing}, and compares \tool to \botium and \charm, which are the state-of-the-art testing tools~\cite{ferdinando2024mutabot,li2022review}. 
We selected Google Dialogflow as the target framework, which is one of the most popular platforms for chatbot design~\cite{abdellatif2021comparison}. Although we cannot make claims about the generalizability of the results to chatbots implemented with other platforms, such as Rasa and Amazon Lex, their similarity suggests that results are unlikely to change drastically among task-based chatbot platforms.  

\vspace{-2mm}
\section{Related Work}\label{sec:related}Chatbots have existed for a long time (e.g., Weizenbaum's creation named Eliza~\cite{weizenbaum1966eliza}), but only recently they have been integrated with AI and natural language processing capabilities~\cite{adamopoulou2020chatbots,sanchez2024automating}. 
Traditional testing methods have proven challenging to apply to this new technology. 
So far, a plethora of methods have focused on performance, usability, design metrics, and guidelines~\cite{laranjo2018conversational,deriu2021survey,canizares2024measuring}. 
On the other hand, few approaches have addressed the functional testing of chatbots~\cite{li2022review}. 

For what concern industrial and open-source testing tools, Amazon released the Alexa Simulator~\cite{alexa-simulator} to test Alexa skills without the need of a device, supporting both text and voice-based interactions. An open-source framework for offline black-box testing of Alexa skills, named Alexa Skill Test Framework~\cite{alexa-test-framework}, has also been provided. The capabilities of the framework include mocking external services and support for audio streams. Both solutions, although highly configurable, are tailored to Alexa. 
Rasa bot builder~\cite{rasa} provides testing features focused on conversational flows and Natural Language Understanding (NLU) models, but test stories must be manually written and maintained. 
Chatbottest~\cite{chatbottest} offers support for improving the quality of chatbot design based on heuristic evaluations, but no actual test generation is supported, and the tool relies on a Chrome extension to evaluate chatbots operating on only a few platforms (e.g., Telegram). Bespoken offers a commercial solution~\cite{bespoken} for conversational AI chatbot testing, using a rich dashboard that integrates utility features, such as voice support and testing reporting. Similarly to Rasa, conversational scenarios in Bespoken must be manually configured. Playwright~\cite{playwright} supports end-to-end testing of chatbots that rely on a Document Object Model (DOM) interface. It employs a recording mechanism to save interactions over the DOM for replication, thus tests have to be recorded manually. 
Unlike these frameworks, \tool provides automated test generation capabilities. 

 \botium~\cite{botium} is an automated quality assurance framework providing both automatic generation and execution of test cases for chatbots, supporting multiple natural language processing engines and chatbot design platforms, including Amazon Lex~\cite{amazon-lex}, Rasa~\cite{rasa} and Dialogflow~\cite{dialogflow}. The test cases generated in \botium are textual conversations between the user and the chatbot. Based on our findings, 
 \botium shows limitations in covering conversations when custom entities and external services are involved.
Further research on mutation testing has shown how test suites generated by \botium struggle to detect numerous mutants affecting conversational properties~\cite{ferdinando2024mutabot,gomez2024mutation}.
To alleviate the cost of manual chatbot testing, 
in 2017 researchers from IBM Research Lab proposed Bottester~\cite{vasconcelos2017bottester}, a tool that simulates user interactions with a chatbot. The tool takes in input a specification of the conversational flows and additional parameters to simulate actual interactions, such as a sleep time to separate each user input. 
\charm~\cite{bravo2020testing} tool has been proposed by Bravo-Santos \emph{et. al} as an extension to \botium for Dialogflow chatbots. \charm enriches \botium conversations using a set of mutations to test chatbot robustness and accuracy.
We empirically compared \tool with \botium and \charm, showing how the \tool's capabilities can improve test effectiveness.

Similarly to \charm, further work addresses input mutation on quality assurance of chatbots. Guichard \emph{et al.}~\cite{guichard2019assessing} investigate the process of automatically paraphrasing user sentences, and they evaluate robustness of text-based conversational agents on the BoTest testing framework~\cite{ruane2018botest}. 
Bozic \emph{et al.} propose an automated approach for functional chatbot testing involving AI planning as foundation for test generation~\cite{bozic2019chatbot}: a plan represents an abstract test case, composed of user's requests as actions to perform and an intent as a goal to achieve, allowing a re-planning phase once stuck. Bozic and Wotawa~\cite{bozic2019testing} further investigate the oracle problem in the chatbot domain. As chatbot output may be difficult to predict, metamorphic testing~\cite{chen2018metamorphic} can be integrated into the testing process, defining metamorphic relations, such as synonym transformations and word removals
. In a subsequent work~\cite{bovzic2022ontology}, they sophisticated the approach introducing
an ontology-based infrastructure. Since employing the ontology requires some effort and knowledge
, the approach can benefit from automated black-box test generator tools, such as \botium, \charm, and \tool. 

\section{Conclusion}\label{sec:conclusion}

The ubiquity of chatbots in human activities, as well as the involvement of advanced technologies for their design, demands suitable quality assurance techniques and tools. So far, only a few approaches have addressed functional testing of task-based chatbots~\cite{bovzic2022ontology,li2022review}. State-of-the-art approaches are limited to the static generation of test cases that cannot always capture the complexity and diversity of questions and responses that are part of conversations, generating test cases that may be fragile or require manual fixtures, providing limited coverage of intents and entities present in a chatbot~\cite{botium,bravo2020testing}. 

In this paper, we have presented \tool, a dynamic test case generation technique for Dialogflow task-based chatbots. \tool leverages \botium test cases to produce test variants capable of exploring scenarios that are not covered by existing tests. The tool records all bot responses, even those depending on external services, and employs setup and teardown operations to clean up the environment, favoring regression testing in a neutral environment. 
Our empirical results show that \tool can generate a higher rate of correct test cases that extensively cover conversations compared to \botium and \charm. 
In future work, our aim is to further refine and expand the capabilities of \tool. 
This will involve conducting additional experiments, such as investigating the adaptability of the tool to other chatbot design platforms (e.g., Rasa), as well as designing more sophisticated test strategies to implement flexible oracles and to cover negative scenarios in the context of LLM-based chatbot testing~\cite{sanchez2024automating}.






\balance

\bibliographystyle{IEEEtran}
\bibliography{IEEEabrv,references}


\end{document}